\documentclass[aps,prb,groupedaddress,twocolumn,10pt,showpacs]{revtex4-1}
\usepackage[utf8]{inputenc}
\usepackage{graphicx}
\usepackage{amsmath,amssymb,color,subfigure,braket,bbm}

\DeclareMathOperator{\Tr}{Tr}

\begin{document}

\title{Local spectroscopies across the superconductor-insulator transition}
\author{Hasan Khan$^{1}$}
\author{Nandini Trivedi$^{1}$}
\affiliation{(1) Department of Physics, The Ohio State University, Columbus, OH  43210, USA}
\date{\today}

\begin{abstract}
We explore the manifestation of quantum fluctuations across the superconductor-insulator transition (SIT) in three different local measurements: the two-particle density of states $P(\mathbf{r},\omega)$, compressibility $\kappa(\mathbf{r})$, and diamagnetic susceptibility $\chi(\mathbf{r})$. We perform Monte Carlo simulations on the 2D quantum Josephson junction array model and present local maps of these quantities as the system is tuned across the SIT. $P(\mathbf{r},\omega)$, obtained using Maximum Entropy analytic continuation techniques, shows strongly diminished zero-energy spectral weight in nearly-insulating islands, that also correlate with regions of suppressed $\kappa(\mathbf{r})$. We investigate the signatures of quantum  fluctuations in the evolution of $\kappa(\mathbf{r})$ and $\chi(\mathbf{r})$ across the SIT, distinct from thermal fluctuations. We discuss the experimental implications of our results for scanning Josephson spectroscopy, compressibility, and scanning SQUID measurements.
\end{abstract}

\maketitle

\section{Introduction}
Since its discovery, superconductivity has long been a rich environment for understanding the quantum nature of our universe. Superconductors themselves are macroscopic condensates of paired electrons that owe their coherence to quantum mechanics. It is no surprise then that superconductivity in disordered two-dimensional thin film materials has proven to be a rewarding playground for studying quantum phase transitions (QPTs).\cite{hebard1990,shahar1992,goldman1998,adams2004,sambandamurthy2004,steiner2005,stewart2007,gantmakher2010} These QPTs are driven by quantum fluctuations between two competing phases of matter at zero temperature tuned by a non-thermal parameter $g$.\cite{sachdev}

For the superconductor-insulator transition (SIT) in thin films, theory\cite{ghosal1998,ghosal2001,bouadim2011} and experiment\cite{sacepe2008,sacepe2011,mondal2011,sherman2012} have both shown that the single-particle gap in the superconductor persists into the insulator. While the local amplitude remains finite and the single-particle density of states shows a robust gap, the coherence peaks become diminished and global superconductivity is lost. This suggests that Cooper pairs do not break up in the insulator, and in fact superconductivity is destroyed by the loss of global phase coherence between pairs on different islands. The SIT, then, is predominantly driven by quantum phase fluctuations. 

We therefore use a bosonic description of a thin film superconductor in terms of a Josephson junction array (JJA) of superconducting islands, the Hamiltonian of which is related to the quantum XY model.\cite{wallin1994,sondhi1997,swanson2014} Here, amplitude fluctuations are ignored and we directly simulate the phase degrees of freedom using quantum Monte Carlo (QMC). This gives us access to quantum phase fluctuations, which have have been experimentally observed in global measurements\cite{crane2007,sherman2015,poran2017} but have only recently been imaged locally using scanning SQUID techniques.\cite{kremen2018}

In this paper we calculate local two-particle quantities such as compressibility, two-particle local density of states (LDOS), and local diamagnetic susceptibility in order to highlight their importance in observing local quantum fluctuations. Local spectroscopies have previously played an important role in identifying a variety of physical phenomena, including the use of scanning tunneling spectroscopy (STM) to map spatial inhomogeneities in high-$\mathrm{T_c}$ cuprates,\cite{howald2001,pan2001,lang2002} local compressibility measurements to observe electron-hole puddles in graphene,\cite{martin2008} and a combination of STM and spin-polarized STM to detect a possible signature of Majorana fermions in ferromagnetic chains on a superconductor.\cite{nadj-perge2014} Here we make predictions for experimental measurements that can be performed using scanning Josephson spectroscopy (SJS)\cite{randeria2016} and compressibility probes, which can be used to visualize quantum phase fluctuations.

The number-phase uncertainty principle that is at the heart of the SIT is also emerging as a paradigmatic model in quantum information. The ``quantum phase slip" qubit describes a state with a well-defined flux in which number fluctuations introduce phase slips. The dual ``Cooper pair box" is a qubit with well defined number of Cooper pairs in which Josephson tunneling creates number fluctuations.\cite{mooij2006} We discuss below how the qubit evolves across the SIT and connect these concepts to our results on spectroscopies.

Our main results are as follows:

\noindent (1) The global two-particle DOS $P(\omega)$ is peaked at $\omega = 0$ in the superconducting phase due to the presence of a Cooper pair condensate. As the SIT is approached, this peak diminishes and the spectrum becomes gapped at the critical point, signaling a transition to a Cooper pair insulator. The global two-particle compressibility $\kappa$ is also finite in the superconducting phase where Cooper pairs are free to tunnel and vanishes at the critical point where they become localized.

\noindent (2) The local compressibility $\kappa(\mathbf{r})$ in a disordered system captures the onset of quantum fluctuations as the SIT is approached. In the superconducting phase, increasing phase fluctuations create pockets of localized Cooper pairs where the compressibility is small. The LDOS $P(\mathbf{r}, \omega)$ in these regions exhibits an $\omega = 0$ peak that is strongly suppressed whereas $P(\mathbf{r}, \omega)$ outside of these pockets resembles the global $P(\omega)$ of a superconductor. As $g$ is increased toward the SIT, fluctuations of $\kappa(\mathbf{r})$ increase mirroring the presence of increasing phase fluctuations. Thermal fluctuations obfuscate the presence of these insulating pockets instead of increasing them in size, providing an easy way to separate quantum fluctuations from thermal fluctuations.

\noindent (3) The local diamagnetic susceptibility $\chi(\mathbf{r})$ also shows increasing fluctuations as the SIT is approached from the superconducting side. As a function of $T/T_c$, the standard deviation of $\chi(\mathbf{r})$ is peaked only in a narrow region around $T_c$ for $g$ deep in the superconducting phase. As $g$ is increased toward the critical point, this peak broadens around $T_c$, indicating that fluctuations of $\chi(\mathbf{r})$ are appearing well-below $T_c$. The fact that these extra fluctuations exist far below $T_c$ provide evidence that they are indeed of quantum origin. 

\section{Model and Methods}
A useful model for understanding quantum fluctuations across the SIT is the 2D JJA model with Hamiltonian

\begin{equation}
\label{eq:ham}
    \hat{H} = \frac{E_C}{2}\sum_i \hat{n}_i^2 - E_J \sum_{\left<ij\right>} \cos(\hat{\theta}_i - \hat{\theta}_j)
\end{equation}
where $\hat{n}_i$ and $\hat{\theta}_i$ are canonically conjugate Cooper pair number and phase operators, respectively, that satisfy the commutation relation $\left[\hat{\theta}_i,\hat{n}_j\right]=i\delta_{ij}$. $E_J$ links phases on nearest neighbor sites via a Josephson coupling while $E_C$ represents the charging energy of Cooper pairs on each site. When $E_J$ is large, the phases align and the system is in a coherent superconducting phase. When the $E_C$ term dominates, the system favors a well-defined number eigenstate, leading to quantum phase fluctuations that destroy the superconducting order and transition the system to a bosonic insulating phase. Thus we use the ratio $g = E_C/E_J$ as a knob to tune the system across a QPT between a superconductor and an insulator. It is important to emphasize that loss of global phase coherence is responsible for destroying superconductivity in our model. We assume that fluctuations of the superconducting amplitude are small and that we are working at temperatures well below the pair-breaking scale $T^\ast$ of the superconducting island.

\begin{figure}[!tb]
\includegraphics[width=0.5\textwidth]{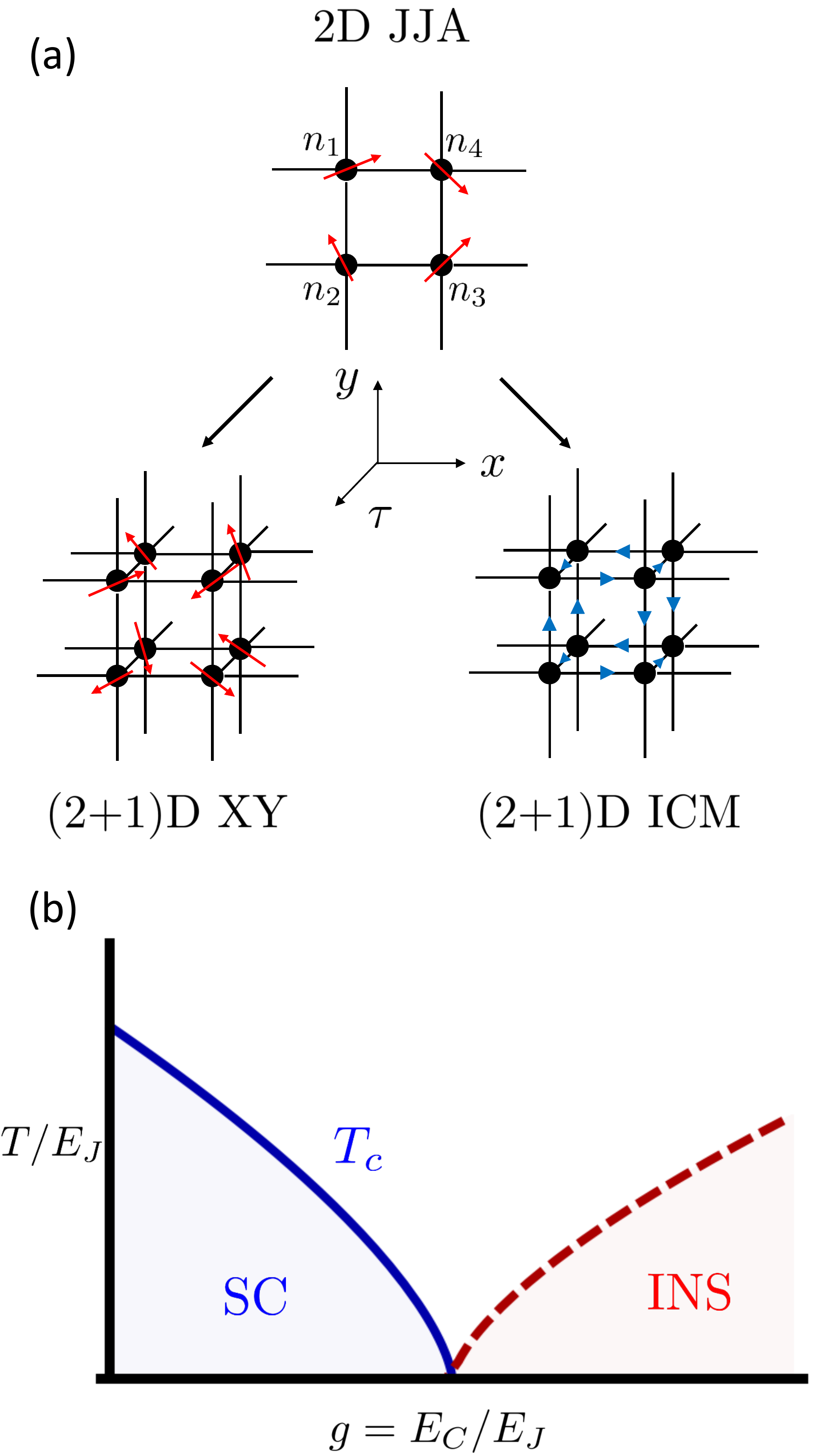}
\caption{(a) We simulate the JJA Hamiltonian by mapping to two separate classical actions: XY and integer current model ICM. (b) Schematic phase diagram of the SIT along the $g$-$T$ plane. At zero temperature, the JJA can be tuned through an SIT at $g = g_c$. At finite temperature, both the superconducting and insulating phases transition to normal states with increasing $T$. In-between there is a quantum critical regime.}
\label{model}
\end{figure}
We simulate this model using QMC two different ways as shown schematically in Fig.~\ref{model}a. In both cases we use a quantum-classical mapping to map the quantum JJA Hamiltonian to a classical action.\cite{wallin1994,sondhi1997} First we map the 2D JJA to a (2+1)D XY model of classical phases, a language that is well suited for calculating the two-particle DOS $P(\omega)$ and compressibility $\kappa$. Simulations of the (2+1)D XY model are performed using a Wolff cluster algorithm\cite{wolff1989} on system sizes $64\times64$ for global calculations and $24\times24$ for local calculations. To calculate diamagnetic susceptibility $\chi$, we instead map the JJA to a (2+1)D integer current model (ICM) because the current basis is more natural for exploring the fluctuations of diamagnetic currents. We simulate the ICM using a worm algorithm\cite{prokofev2001} on a $64\times64$ lattice for both global and local calculations. See Appendix A for more details on these mappings.

It is important to note that simulations calculating local quantities require a small amount of disorder to be introduced in order to create structure. Without disorder, local structure is washed out by Monte Carlo averaging. We introduce a small amount of disorder in the spatial bonds of each model by randomly removing a fraction $p = 0.1$ of Josephson couplings $E_J$ throughout the lattice. This creates regions where insulating sites can nucleate and be detected by our local probes. There have previously been studies on disorder-tuned SITs,\cite{bouadim2011,swanson2014} however we emphasize that the disorder in our model is static and the fluctuations we see are ultimately caused by tuning $g$, not disorder.

Both the global $P(\omega)$ and local $P(\mathbf{r},\omega)$ are calculated by analytically continuing imaginary time data to real frequencies using the Maximum Entropy Method (MEM).~\cite{gubernatis1991,sandvik1998} The procedure is delicate and we have performed extensive tests, including checking sum rules, to ensure the validity of our results (see Appendix B for more information).

\section{Results}
\subsection{Bosonic spectral function}
\begin{figure*}[!tb]
\includegraphics[width=\textwidth]{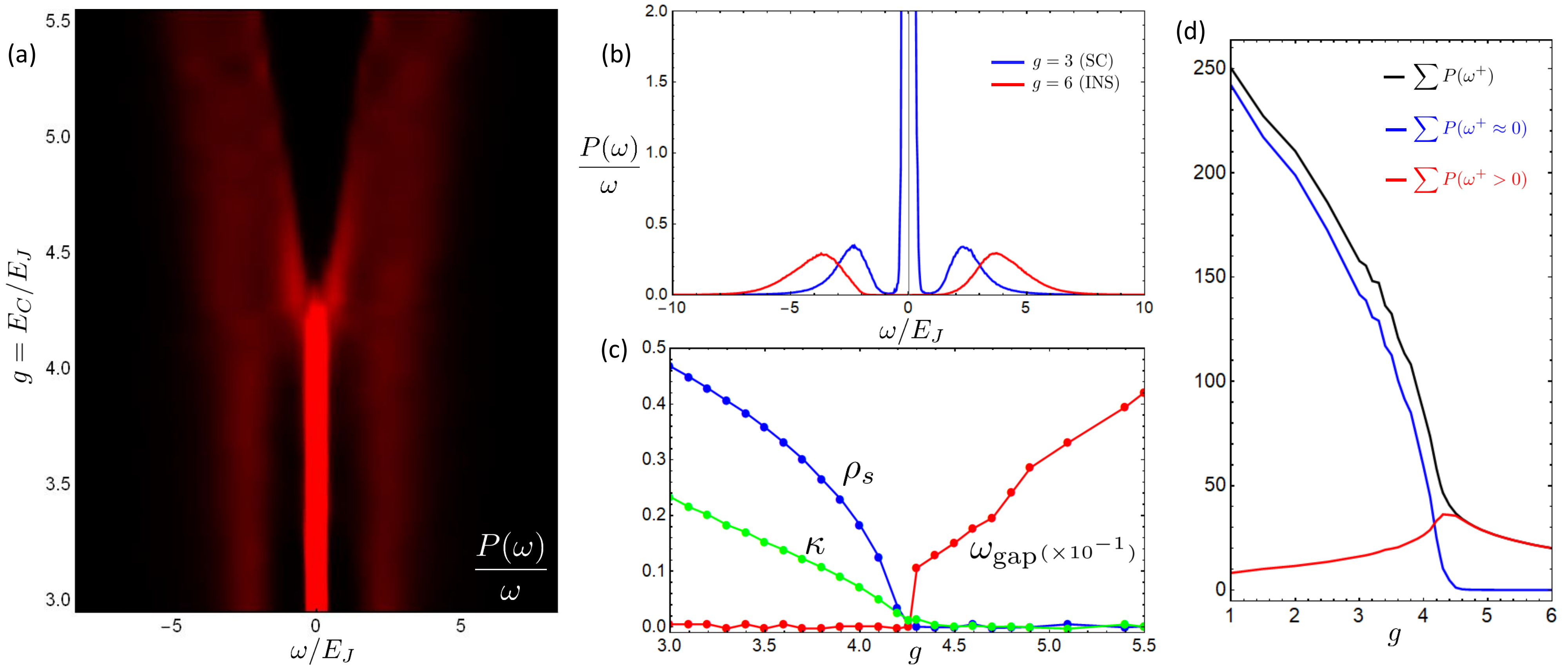}
\caption{Global density of states and energy scales across the SIT: (a) Global two-particle DOS $P(\omega)/\omega$ obtained from analytic continuation of the imaginary time Green's function $G(\tau)$ shown as a function of $g$. For $g < g_c \sim 4.26$ the DOS shows a zero energy peak corresponding to the Cooper pair condensate. As $g$ increases, this weight shifts toward finite energy modes and for $g > g_c$ the system forms a gap characterized by the energy scale $\omega_{\mathrm{gap}}$, signaling the transition to an insulating state. (b) Cuts of $P(\omega)/\omega$ plotted for $g = 3$ (superconducting) and $g = 6$ (insulating) highlighting the difference in two-particle spectra between the two phases. (c) Energy scales near the SIT. The superfluid stiffness $\rho_s$ and compressibility $\kappa$ are finite in the superconducting phase but go soft at $g_c$. Similarly, on the insulating side, the two-particle gap scale $\omega_{\mathrm{gap}}$ is finite and approaches zero as $g_c$ is approached. Note that the error bars are the size of the data points here. (d) We show how the spectral weight in $P(\omega)$ shifts from $\omega = 0$ to finite energy modes with increasing $g$. At small $g$, most of the weight is centered around zero energy, however as $g$ increases past the SIT this weight decreases to zero and increasingly shifts to finite energy states.}
\label{global}
\end{figure*}
The bosonic Green's function in imaginary time is given by $G(\mathbf{r}, \mathbf{r}'; \tau) = \left< \hat{a}^\dagger (\mathbf{r}',\tau) \hat{a}(\mathbf{r},0) \right>$ where $\hat{a}$, $\hat{a}^\dagger$ are Cooper pair raising and lowering operators. In the language of the JJA we can rewrite\cite{fisher1989} the raising operator in terms of its amplitude and phase as $\hat{a}^\dagger (\mathbf{r},\tau) = \sqrt{\hat{n}(\mathbf{r},\tau)}e^{i\hat{\theta}(\mathbf{r},\tau)}$, but since we are ignoring on-site amplitude fluctuations we can write the Green's function purely in terms of the phase variable as
\begin{equation}
\label{eq:green}
    G(\mathbf{r}, \mathbf{r}'; \tau) = \left< e^{i\left(\hat{\theta}(\mathbf{r}',\tau) - \hat{\theta}(\mathbf{r},0)\right)} \right>
\end{equation}
which is a spin-spin correlation function in the classical XY representation.

The real frequency spectral function $P(\mathbf{k},\omega)$ is the imaginary part of the corresponding real frequency Green's function $G(\mathbf{k},\omega)$

\begin{equation}
    P(\mathbf{k},\omega) = - \frac{1}{\pi}\mathrm{Im} G(\mathbf{k},\omega).
\end{equation}
However, since we are working with imaginary time in our QMC, we need a way to analytically continue $G(\mathbf{k},\tau)$ to a real frequency spectral function. This leads to the following relation between $G(\mathbf{k},\tau)$ and $P(\mathbf{k},\omega)$

\begin{equation}
\label{eq:laplace}
    G(\mathbf{k},\tau) = \int_{-\infty}^\infty \frac{d\omega}{\pi} \frac{e^{-\tau\omega}}{1-e^{-\beta\omega}} P(\mathbf{k},\omega).
\end{equation}
Solving this equation for $P(\mathbf{k},\omega)$ amounts to inverting a Laplace transform. However, performing this procedure on QMC data of $G(\mathbf{k},\tau)$ with error bars is non-trivial and requires the use of numerical analytic continuation techniques. In our work, we use MEM to obtain $P(\mathbf{k},\omega)$ from $G(\mathbf{k},\tau)$ and have validated our results by checking relevant sum rules.

We are particularly interested in the DOS which is the sum over momentum of the spectral function
\begin{equation}
    P(\omega) = \sum_{\mathbf{k}} P(\mathbf{k},\omega) = A(|\mathbf{r}-\mathbf{r}'| = 0,\omega)
\end{equation}
This amounts to performing MEM on $G(|\mathbf{r}-\mathbf{r}'|=0,\tau)$ data. In Fig.~\ref{global}a we plot $P(\omega)/\omega$ as a function of $g$ across the SIT. Since the form of (\ref{eq:laplace}) requires $A(-\omega) < 0$, we plot $P(\omega)/\omega$ to obtain a quantity that more closely resembles an experimentally measured DOS. We see that $P(\omega)/\omega$ is strongly peaked at $\omega = 0$ in the superconducting phase, corresponding to the existence of a Cooper pair condensate. As the SIT is approached, spectral weight shifts from the central peak to the finite energy modes on either side of $\omega$ until the system forms a gap for $g>g_c$. We plot the shift of this spectral weight from zero energy to finite energy modes in Fig. \ref{global}d. In Fig. \ref{global}c we plot the size of this gap $\omega_{\mathrm{gap}}$ as a function of $g$ along with other energy scales including the superfluid stiffness $\rho_s$ and the compressibility $\kappa$ (described in the next section). As we expect, $\rho_s$, $\kappa$, and $\omega_{\mathrm{gap}}$ go soft at $g_c$ from their respective sides of the transition.

\subsection{Compressibility}
The global compressibility $\kappa$ is a quantity that characterizes fluctuations of the Cooper pair number density operator $\hat{n}$. We can define the average current along a generic spacetime bond $b$ in the XY representation as 

\begin{equation}
    \braket{j_b} = -\frac{\partial \ln Z}{\partial A_b} = \braket{K_b\sin(\partial_b \theta - A_b)}
\end{equation}
where $Z = \Tr e^{-\beta \hat{H}}$ is the partition function, $A_b$ is the element of an externally applied vector potential along bond $b$, and $K_b$ is the coupling constant along that bond. We can then identify the average density $\braket{n}$ with the current along \textit{temporal} bonds $\braket{j_\tau}$.\cite{wallin1994} 

The generalized electromagnetic response tensor 

\begin{equation}
    \Upsilon_{bb'} = \frac{\partial \braket{j_b}}{\partial A_{b'}}
\end{equation}
describes the response of a current $j_b$ along a spacetime bond $b$ to an externally applied vector potential $A_{b'}$ along a bond $b'$. While the superfluid stiffness $\rho_s$ can be obtained from the static, transverse long wavelength limit of the spatial response function $\Upsilon_{xx}$, the compressibility is given by the static long wavelength limit of the temporal response function $\Upsilon_{\tau\tau}$~\cite{kubo2003}

\begin{align}
    \Upsilon_{\tau\tau}(\mathbf{r},\mathbf{r}';\tau,\tau') &=\frac{\partial \braket{j_\tau (\mathbf{r},\tau)}}{\partial A_\tau(\mathbf{r}',\tau')}\\ 
    \label{eq:comp}
    &= \left<-k_\tau(\mathbf{r},\tau)\right>\delta(\mathbf{r},\tau) - \Lambda_{\tau\tau}(\mathbf{r},\tau)
\end{align}
where $\left<-k_\tau(\mathbf{r},\tau)\right> = \left<K_\tau \cos(\partial_\tau\theta(\mathbf{r},\tau))\right>$ and the temporal current-current correlator $\Lambda_{\tau\tau}(\mathbf{r},\tau) = \left<K_\tau^2 \sin(\partial_\tau\theta(\mathbf{r},\tau))\sin(\partial_\tau\theta(0,0))\right>$. Note that since we are performing linear response we take the limit $A_\tau \rightarrow 0$, and we make use of translational symmetry in the second line.

The compressibility $\kappa$ is then the static, long wavelength limit of $\Upsilon_{\tau\tau}(\mathbf{r},\tau)$
\begin{equation}
    \kappa = \lim_{\mathbf{k}\rightarrow 0} \Upsilon_{\tau\tau}(\mathbf{k},i\omega_n=0).
\end{equation}
Note that this amounts to calculating the response of the pair number density $\braket{n} \sim \braket{j_\tau}$ with respect to an externally applied electric potential $\phi \sim A_\tau$, which is the usual definition of compressibility.

In Fig. \ref{global}c we show $\kappa$ as a function of $g$. $\kappa$ is finite in the superconducting phase where Cooper pairs are phase coherent and are able to tunnel across the system. $\kappa$ decreases as $g_c$ is approached and vanishes in the Mott insulating phase where Cooper pairs become localized by phase fluctuations.

\subsection{Local quantities}
We next turn our attention to calculations of the LDOS $P(\mathbf{r},\omega)$ and local compressibility $\kappa(\mathbf{r})$. $P(\mathbf{r},\omega)$ is related to the local Green's function (\ref{eq:green}) at $\mathbf{r}=\mathbf{r}'$ by inverting (\ref{eq:laplace}) once again

\begin{equation}
    G(\mathbf{r},\tau) = \int_{-\infty}^\infty \frac{d\omega}{\pi} \frac{e^{-\tau\omega}}{1-e^{-\beta\omega}} P(\mathbf{r},\omega).
\end{equation}
The local compressibility $\kappa(\mathbf{r})$ is given by the local response function $\Upsilon_{\tau\tau}(\mathbf{r},\tau)$ in (\ref{eq:comp})
\begin{align}
   \kappa(\mathbf{r}) &= \Upsilon_{\tau\tau}(\mathbf{r},i\omega_n=0) = \lim_{A_\tau\rightarrow 0} \frac{1}{\beta}\sum_{\mathbf{r}',\tau,\tau'} \frac{\partial j_\tau(\mathbf{r},\tau)}{\partial A_\tau(\mathbf{r}',\tau')} \\
   &= \frac{1}{\beta}\sum_{\mathbf{r}',\tau,\tau'}\left(\left<-k_\tau(\mathbf{r},\tau)\right>\delta(\mathbf{r}',\tau') - \Lambda_{\tau\tau}(\mathbf{r},\mathbf{r}',\tau,\tau')\right).
\end{align}
\begin{figure*}[!tb]
\includegraphics[width=\textwidth]{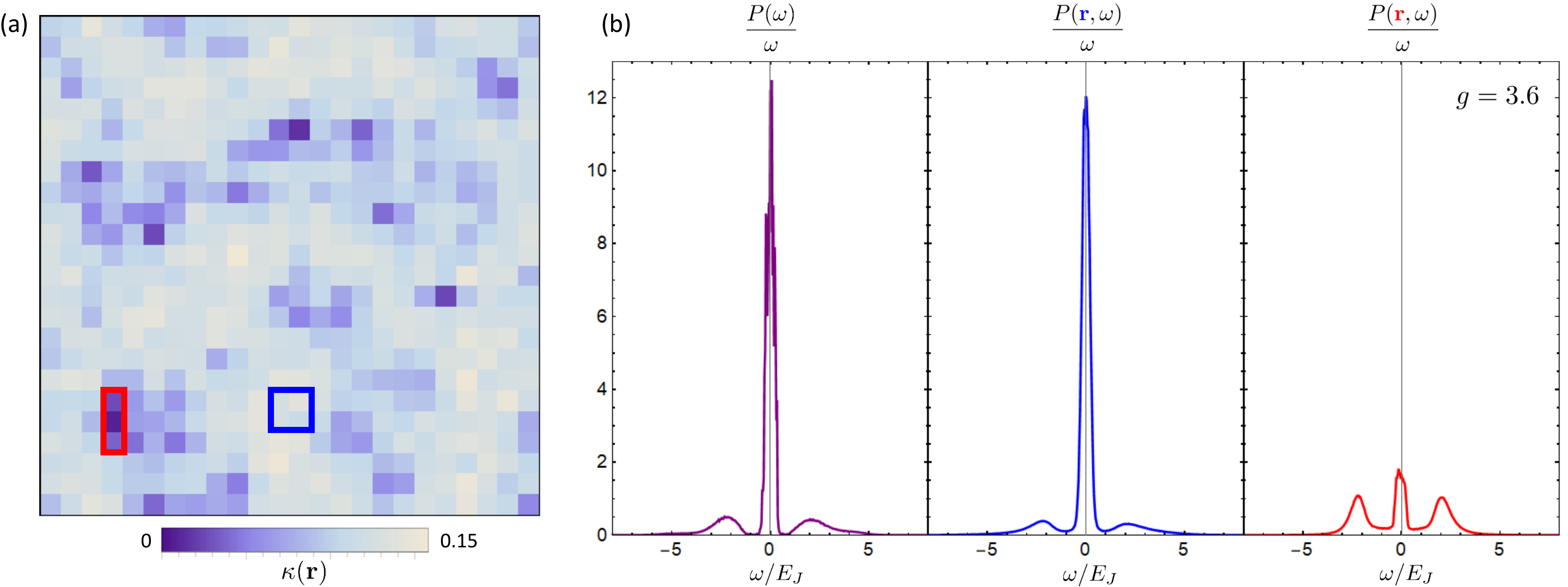}
\caption{Local compressibility and LDOS in a disordered system near the SIT. (a) Map of the local compressibility $\kappa(\mathbf{r})$ on a $24\times24$ lattice at $g = 3.6$, near the SIT. We introduce a small amount of bond disorder ($p=0.1$) to produce local structure in our QMC data. We see that while the majority of system has finite $\kappa$ (superconducting), the local $\kappa(\mathbf{r})$ map picks out large dark regions with $\kappa$ near zero (insulating), as we would expect for a system exhibiting strong quantum fluctuations. This is reflected in the two-particle LDOS $P(\mathbf{r}, \omega)/\omega$ shown in (b). In the compressible region highlighted in blue, we see that $P(\mathbf{r}, \omega)/\omega$ has a peak at $\omega = 0$ characteristic of a superconductor. On the other hand, the incompressible region highlighted in red exhibits a peak that is highly suppressed at $\omega = 0$, indicating that this region is approaching an insulating regime. The global $A( \omega)/\omega$ is shown in purple for comparison.}
\label{localDOS}
\end{figure*}
In order to extract local structure from our QMC simulations, we break translational symmetry by introducing a small fraction of bond disorder. In Fig. \ref{localDOS}a we show a local map of $\kappa(\mathbf{r})$ at $g=3.6$, near the SIT. We see that the system forms puddles where $\kappa(\mathbf{r})$ is significantly suppressed. These incompressible regions are locations where a large density of bonds have been cut, resulting in the formation of insulating islands. In Fig. \ref{localDOS}b we plot the corresponding $P(\mathbf{r},\omega)$ in two representative regions. We see a strong spatial correlation between the strength of $\kappa(\mathbf{r})$ and the distribution of low-energy spectral weight. The superconducting region highlighted in blue is highly compressible and has most of its spectral weight peaked strongly around $\omega = 0$, reflecting the strength of the superconducting condensate. The behavior of $P(\mathbf{r},\omega)$ in this region matches that of the global $P(\omega)$ shown in purple, which reflects that of a globally phase-coherent superconductor. On the other hand, in the region highlighted in red with small compressibility, we see that the $\omega = 0$ peak in $P(\mathbf{r},\omega)$ is highly suppressed . Here we can see evidence of a two-particle gap beginning to form as spectral weight shifts from $\omega = 0$ to finite energy modes, indicative of an emerging Cooper pair insulator.

\begin{figure*}[!tb]
\includegraphics[width=\textwidth]{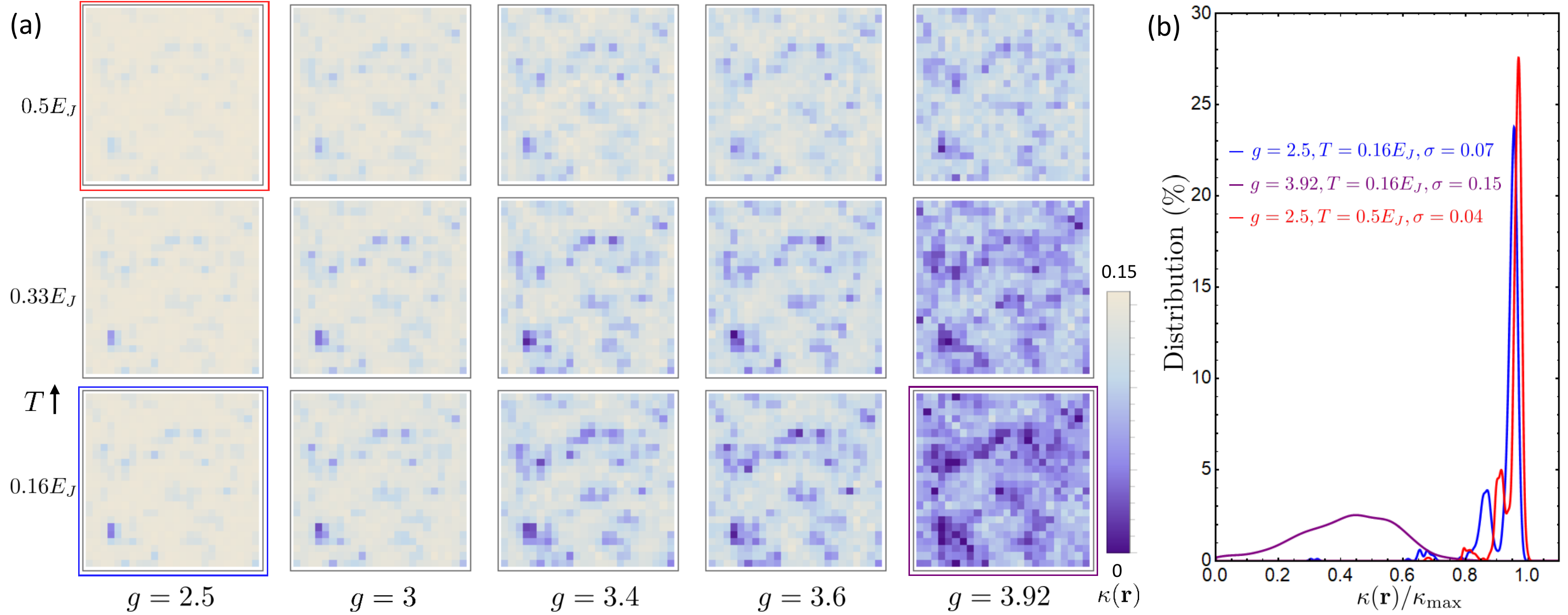}
\caption{(a) Maps of the local compressibility $\kappa(\mathbf{r})$ on a $24\times24$ lattice as a function of $g$ and temperature $T$ on the superconducting side of the transition. We see that as the transition is approached with increasing $g$, fluctuations of the compressibility also increase. Interestingly, with increasing $T$ we see that fluctuations in $\kappa(\mathbf{r})$ actually become smoothed out due to increasing number fluctuations. Since the behavior of $\kappa(\mathbf{r})$ as we evolve tuning $g$ or $T$ is different, we can separate the effects of thermal fluctuations from quantum fluctuations. (b) Distributions of $\kappa(\mathbf{r})$ for various maps. We see that as $g$ increases for fixed $T$, the distribution of $\kappa(\mathbf{r})$ broadens significantly and the standard deviation $\sigma$ increases. However, with increasing $T$ and fixed $g$, the distribution only changes slightly and actually becomes narrower.}
\label{localKappa} 
\end{figure*}

It is important to emphasize that the emergence of insulating islands shown in Fig. \ref{localDOS}a is caused by quantum phase fluctuations due to proximity to a quantum critical point. To illustrate this, in Fig. \ref{localKappa}a we also plot $\kappa(\mathbf{r})$ as a function of $g$ and temperature $T$. As $g$ increases toward the SIT, fluctuations in $\kappa(\mathbf{r})$ increase, leading to an increase in the size and prevalence of incompressible islands. Interestingly, as $T$ increases the fluctuations in $\kappa(\mathbf{r})$ become smoothed out and the insulating islands become smaller. This is due to the fact that $\kappa$ is sensitive specifically to number fluctuations. While quantum number fluctuations are expected to decrease as $g$ increases, \textit{thermal} number fluctuations increase as $T$ increases due to higher energy number states becoming available. This is clearly seen in Fig. \ref{localKappa}b, where the distribution of $\kappa(\mathbf{r})$ broadens significantly with increasing $g$, but narrows slightly with increasing $T$. This difference in behavior as $\kappa(\mathbf{r})$ evolves with $g$ and $T$ provides a way to distinguish quantum fluctuations from thermal fluctuations. We propose that an experiment that measures $\kappa(\mathbf{r})$ across the SIT will be able to directly image the presence of quantum fluctuations. The fact that these fluctuations are in fact \textit{quantum} phase fluctuations is further confirmed by our results on the diamagnetic susceptibility.

\subsection{Diamagnetic susceptibility}
\begin{figure*}[!tb]
\includegraphics[width=\textwidth]{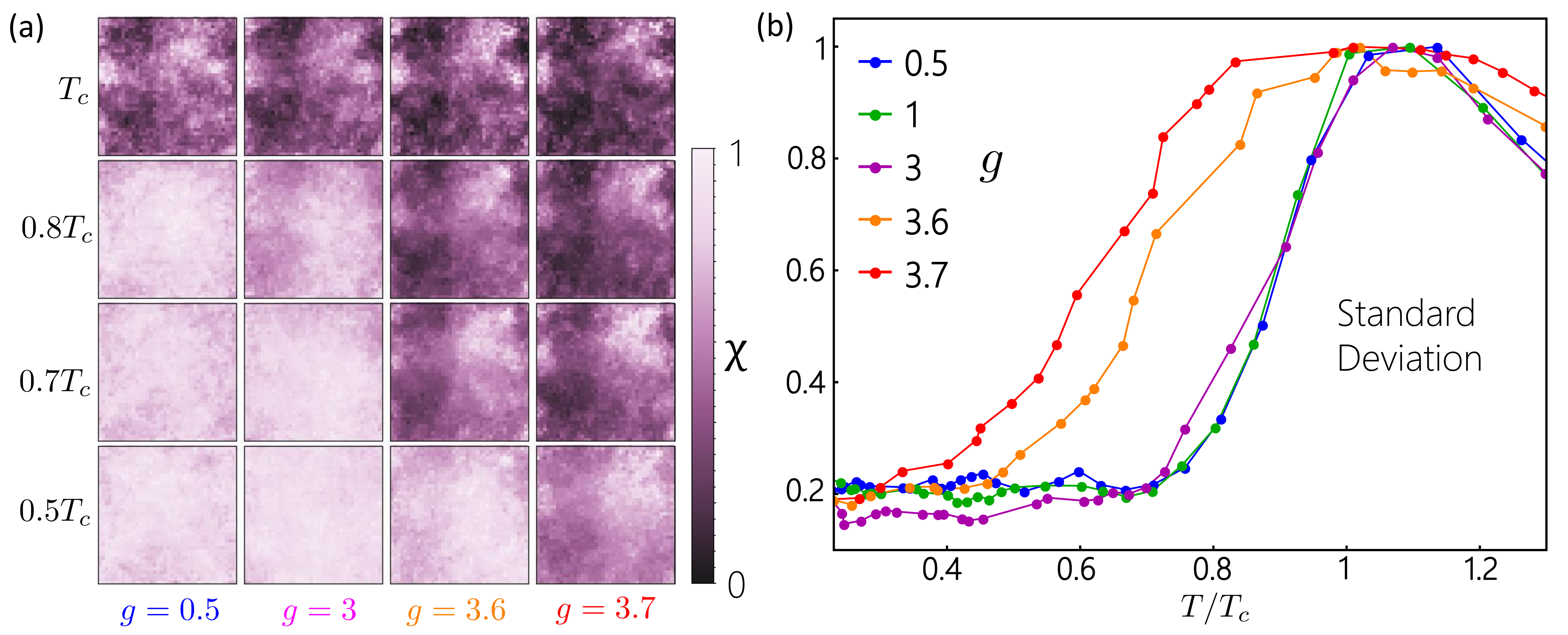}
\caption{(a) Local maps of the diamagnetic susceptibility $\chi(\mathbf{r})$ obtained on a $64\times64$ lattice as a function of $g$ and $T$ using the ICM representation. We see that for small $g$, $\chi(\mathbf{r})$ is large and uniform until the system approaches $T_c$, as expected of thermal fluctuations. However, as $g$ is increased toward the SIT, fluctuations in $\chi(\mathbf{r})$ begin to appear well below $T_c$, suggesting that these additional fluctuations are quantum in nature. In (b) we plot the standard deviation of $\chi(\mathbf{r})$ for each value of $g$ as a function of temperature. We see that while the standard deviation is always peaked around $T_c$, this peak broadens as the SIT is approached, pointing to the increasing importance of quantum phase fluctuations in this regime.}
\label{dia}
\end{figure*}
Phase fluctuations increase both as a function of temperature and a function of $g$. The question becomes how to separate thermal phase fluctuations from quantum phase fluctuations. A well-known property of a superconductor is the fact that it generates diamagnetic supercurrents in the presence of a magnetic field. In general the magnetization generated by an external field is related to the free energy by

\begin{equation}
    \braket{M} = \frac{\partial (T\log Z)}{\partial B}
\end{equation}
where $-T\log Z$ is the free energy in terms of the partition function $Z$ and $B$ is an external applied magnetic field. We can calculate the corresponding local diamagnetic susceptibility $\chi(\mathbf{r})$, which is sensitive to phase fluctuations, from linear response using a Kubo formula

\begin{equation}
    \chi(\mathbf{r}) = -\frac{\partial \braket{M(\mathbf{r})}}{\partial B} = \left<M(\mathbf{r})M\right>.
\end{equation}
The local diagmagnetic susceptibility is the response of a \textit{local} induced magnetization $M(\mathbf{r})$ to a \textit{global} applied magnetic field $B$. This amounts to calculating the correlator between the local magnetization $M(\mathbf{r})$ and the global magnetization $M$. To obtain $\chi(\mathbf{r})$, we perform QMC simulations in the dual ICM representation of the quantum JJA model. This representation is more suited to calculating quantities involving the supercurrent since the QMC configurations themselves are given in terms of integer currents (see Appendix A2).

\begin{figure*}[!htb]
\includegraphics[width=\textwidth]{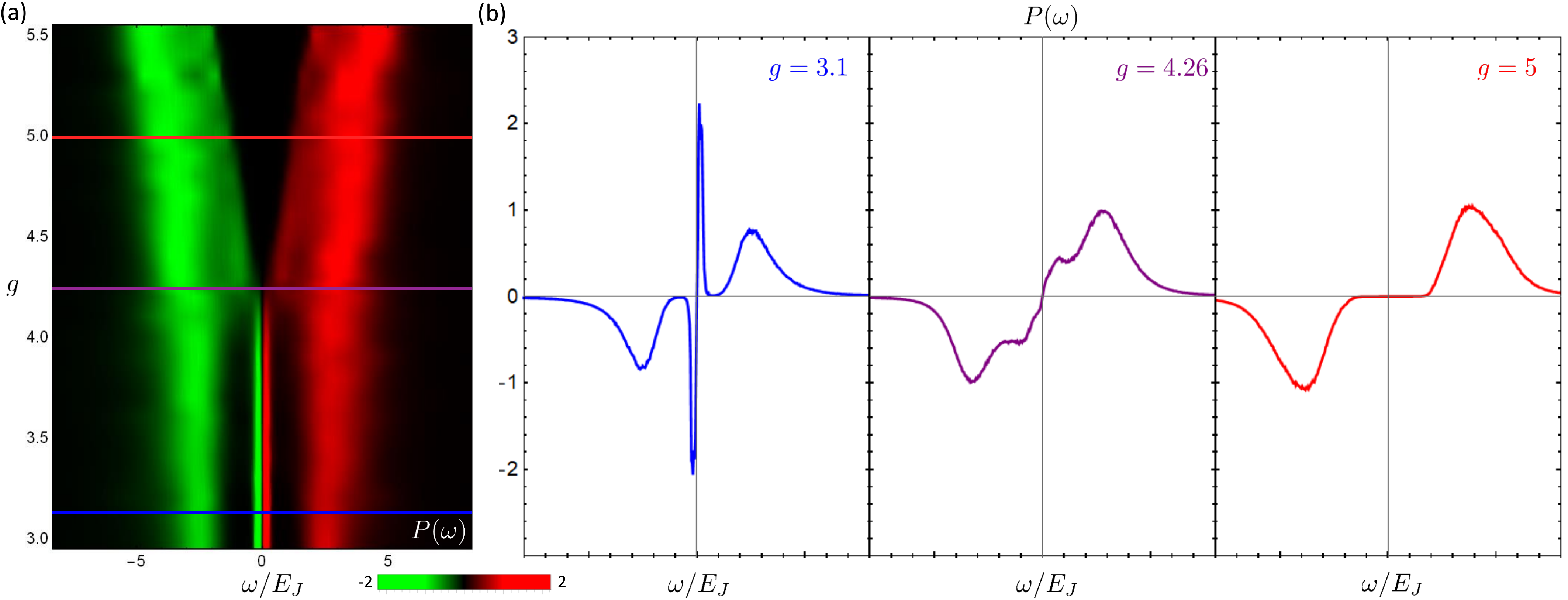}
\caption{(a) Evolution of the Cooper pair spectral function $P(\omega)$ across the SIT tuned by $g=E_c/E_J$, the ratio of charging energy to the Josephson energy. (b) The first cut shows a ``Quantum phase slip" qubit whose behavior is consistent with a finite current at zero voltage, if we interpret the y-axis as the current and the x-axis as the voltage. In the second cut, we approach the transition described by a finite slope around $\omega=0$, consistent with a finite resistance. The final cut describes a ``Cooper pair box" qubit whose behavior is consistent with zero current until a critical voltage is reached.}
\label{IV}
\end{figure*}

In the language of the ICM, we can write the total magnetization, assuming a uniform $B$-field in the $\hat{z}$-direction, as
\begin{equation}
    \braket{M} = \frac{1}{2\beta}\sum_{\braket{ij},\tau} (x_i y_j - x_j y_i)j_{ij}^\tau
\end{equation}
where $j_{ij}^\tau$ is an integer current on a spatial bond connecting sites $i$ and $j$ on timeslice $\tau$. The spatial pattern that $j_{ij}^\tau$ is summed along comes from the corresponding vector potential of the uniform $B$-field, $\mathbf{A} = \frac{B}{2}(-x,y,0)$. $M(\mathbf{r})$ is calculated from the discrete analog of the magnetization current $\mathbf{j}(\mathbf{r}) = \Delta\times\mathbf{M}(\mathbf{r})$ where $\mathbf{j}$ are the integer currents. Inverting this for $M_z(\mathbf{r}) = M(\mathbf{r})$ gives us
\begin{equation}
    M(\mathbf{r}) = \int d\mathbf{r}' \cdot (\hat{z} \times \mathbf{j}(\mathbf{r}')).
\end{equation}

In Fig. \ref{dia}a we show local maps of $\chi(\mathbf{r})$ as a function of $g$ and $T/T_c$. Here, $T_c$ is the temperature at which the superconductor transitions to a normal state but still contains pairs; it should not be confused with the pair-breaking transition, that occurs at a much higher temperature. As expected, we observe that fluctuations of $\chi(\mathbf{r})$ increase with $T/T_c$ due to thermal fluctuations. Similarly, fluctuations of $\chi(\mathbf{r})$ also increase with increasing $g$. In both cases we see pockets with near-zero susceptibility beginning to form as phase fluctuations destroy the ability for coherent supercurrents to exist. However, interestingly, as $g$ increases the fluctuations in $\chi(\mathbf{r})$ appear at lower values of $T/T_c$. The fact that fluctuations appear well below $T_c$ when the system is near a quantum critical point suggests that the fluctuations with increasing $g$ are predominantly caused by \textit{quantum} phase fluctuations.

This shows up as a broadening of the standard deviation of $\chi(\mathbf{r})$ across $T/T_c$ as shown in Fig. \ref{dia}b. For small $g$, the standard deviation is peaked mainly around $T_c$, suggesting that only thermal fluctuations are relevant in this regime. However as $g$ increases the temperature range where standard deviation is large broadens to include temperatures well below $T_c$ reflecting the importance of quantum fluctuations. This broadening of the temperature range of $\chi(\mathbf{r})$ fluctuations was recently observed in scanning SQUID experiments of the thin film superconductor NbTiN.\cite{kremen2018} In the experiment, a scanning SQUID was used to directly image the local $\chi(\mathbf{r})$ and the corresponding standard deviation as a function of temperature and film thickness was found to qualitatively match that of theory. Importantly, this was the first time quantum fluctuations were directly imaged in an experiment.

\section{Conclusion and Outlook}
We have presented three different calculations of local two-particle response functions across the SIT: density of states $P(\mathbf{r},\omega)$, compressibility $\kappa(\mathbf{r})$, and diamagnetic susceptibility $\chi(\mathbf{r})$. We have shown how these quantities can be used to probe the local structure of fluctuations across the SIT. Particularly, for both $\chi(\mathbf{r})$ and $\kappa(\mathbf{r})$ we see an increase in local quantum fluctuations as the system approaches the critical point from the superconducting side, independent of thermal fluctuations

Our aim is to connect these calculations to spectroscopic measurements that can be performed in an experiment. We have already discussed how $\chi(\mathbf{r})$ can be measured using scanning SQUID techniques. We next turn our attention to measurements of $P(\omega)$ and $\kappa(\mathbf{r})$. It has previously been shown that it is possible capture the local structure of the superconducting order parameter using a combination of scanning tunneling spectroscopy (STM) and scanning Josephson spectroscopy (SJS) using a superconducting Pb tip.\cite{randeria2016} There, a suppression of the zero-energy peak measured in the SJS conductance was found on impurity sites, similar to our results on disorder sites of the JJA. We propose an experiment that uses SJS in conjunction with local compressibility measurements\cite{martin2008} to map out the evolution of quantum fluctuations across the SIT as we have done in Fig.~\ref{localDOS} and \ref{localKappa}.

We are also interested in connecting with recent developments in quantum information and quantum computing. As emphasized earlier, the essence of the SIT is the number-phase uncertainty principle. This can be used to define two types of dual qubits based on whether the phase dominates with quantum phase slips (QPS) disrupting that order (``QPS" qubit) on the SC site or whether the number of Cooper pairs is well-defined with Cooper pair tunneling disrupting the order (``Cooper pair box" qubit) on the insulating side.\cite{mooij2006} The behavior of the Cooper pair spectral function in Fig.~\ref{IV} tracks the evolution of the qubit across the SIT .

\noindent {\it QPS qubit:} 
The QPS qubit is dominated by the inductive energy $E_J=\Phi_0^2/{(2L)}$, where $L$ is the inductance of the loop and $\Phi_0=h/{(2e)}$ is the SC flux quantum. The charging term mixes states with different fluxoid number $f=\Phi/\Phi_0$ where $\Phi$ is the flux through the loop and lifts the degeneracy at half-integer values of $f$. A current biased Josephson junction can be modeled as the dynamics of the phase in a slanted washboard potential. The phase is trapped in one of the minima yielding a zero voltage state, a superconductor, until the current exceeds a critical value. 

\noindent {\it Cooper pair box qubit:} 
In the dual regime, the insulator has a fixed number of Cooper pairs on each island and is dominated by the charging scale $E_C=(2e)^2/{(2C)}$ where $C$ is the capacitance of the island. Josephson tunneling mixes states with $n$ and $n+1$ Cooper pairs on an island and lifts the degeneracy at half integer values. In the voltage-biased configuration, charge is trapped in a potential minimum resulting in a zero current state, an insulator, for voltages below a critical value.

Once we understand how a qubit behaves in different regimes, it in fact becomes a device to measure and quantify the degree of fluctuations, both thermal and quantum, across QPTs. We expect these ideas will motivate developments in quantum measurement.

\section*{Acknowledgements}
H.K. would like to acknowledge support from the Israel US bi-national foundation grant no. 2014325. N.T. acknowledges support from the DOE-BES grant DE-FG02-
07ER46423. We would like to thank Yen Lee Loh for helpful discussions.

\section*{Appendix A: Quantum-Classical Mapping}
Here we summarize the mapping from the quantum JJA Hamiltonian (\ref{eq:ham}) to two different classical actions for use in Monte Carlo simulations, the XY model $S_{\mathrm{XY}}$ and the integer current model $S_{\mathrm{ICM}}$. This procedure is carried out as detailed in Wallin et. al.\cite{wallin1994}

\subsection*{Appendix A1: Mapping to Classical XY Model}
The imaginary time path integral formalism allows us to map the JJA Hamiltonian to a corresponding classical one which can be easily simulated in Monte Carlo. The process begins with taking the partition function of the quantum system

\begin{equation}
\label{eq:part}
Z = \Tr e^{-\beta\hat{H}}
\end{equation}
and breaking the inverse temperature $\beta$ into $M$ imaginary timeslices of width $\delta\tau$, $\beta = M\delta\tau$. The partition function, written in the phase basis, then becomes

\begin{align}
Z &= \int\left[\frac{1}{\left(2\pi\right)^{L^d}}\prod_{\mathbf{r}}d\theta_\mathbf{r}\right]\braket{\theta|e^{-M\delta\tau\hat{H}}|\theta}\\
&= \int \mathcal{D}\theta \braket{\theta|e^{-\delta\tau\hat{H}}\ldots e^{-\delta\tau\hat{H}}e^{-\delta\tau\hat{H}}|\theta}
\label{eq:part2}
\end{align}
where the product in $\mathcal{D}\theta$ runs over all the sites of a lattice of size $L^d$ and the states $\ket{\theta} = \prod_\mathbf{r}\ket{\theta_\mathbf{r}}$. We can now insert $M-1$ complete sets of states 

\begin{equation}
\mathbbm{1}_\tau = \int\left[\frac{1}{\left(2\pi\right)^{L^d}}\prod_{\mathbf{r}}d\theta_\mathbf{r}^\tau\right]\ket{\theta^\tau}\bra{\theta^\tau}
\end{equation}
between each exponential in~\eqref{eq:part2}, using the imaginary timeslice index $\tau$ to keep track of each identity.

In this form the partition function becomes

\begin{align}
Z &= \int\mathcal{D}\theta\braket{\theta^0|e^{-\delta\tau\hat{H}}|\theta^{M-1}}\ldots\braket{\theta^2|e^{-\delta\tau\hat{H}}|\theta^1}\braket{\theta^1|e^{-\delta\tau\hat{H}}|\theta^0}\\
\label{eq:part3}
&= \int\mathcal{D}\theta\prod_{\tau=0}^{M-1}\braket{\theta^{\tau+1}|e^{-\delta\tau\hat{H}}|\theta^\tau}\delta\left(\theta^M - \theta^0\right)
\end{align}
where we have absorbed the differential and prefactors of the identity into $\mathcal{D}\theta$.

We can now make the interpretation that the partition function~\eqref{eq:part3} is a sum over paths of phase configurations $\ket{\theta}$ propagating across imaginary time, which is periodic in $\beta$. Thus far everything we have done has been exact. The form of~\eqref{eq:part3} allows us to make an important simplifying approximation. If we choose a suitably large number of timeslices $M$ such that $\delta\tau$ is small, we can perform a Suzuki-Trotter decomposition\cite{suzuki1993} on the exponential $e^{-\delta\tau\hat{H}} = e^{-\delta\tau\left(\hat{T}+\hat{V}\right)} = e^{-\delta\tau\hat{T}}e^{-\delta\tau\hat{V}} + \mathcal{O}\left(\delta\tau^2\right)$ and drop terms of $\mathcal{O}\left(\delta\tau^2\right)$

\begin{equation}
Z \approx \int\mathcal{D}\theta\prod_{\tau=0}^{M-1}\braket{\theta^{\tau+1}|e^{-\delta\tau\hat{T}}e^{-\delta\tau\hat{V}}|\theta^\tau}\delta\left(\theta^M - \theta^0\right).
\end{equation}
Here, $\hat{T}$ and $\hat{V}$ represent the $E_C$ and $E_J$ terms in $\hat{H}$ respectively. Since $\ket{\theta}$ are eigenstates of $\hat{V}$, the second exponential pulls out the familiar classical XY cosine term 

\begin{align}
Z \approx \int\mathcal{D}\theta e^{\delta\tau E_J\sum_{\braket{\mathbf{r}\mathbf{r}'},\tau}\cos\left(\theta_\mathbf{r}^\tau-\theta_{\mathbf{r}'}^\tau\right)}\prod_{\tau=0}^{M-1}T_\tau\delta\left(\theta^M - \theta^0\right)
\label{eq:part4}
\end{align}
where what remains is to evaluate the $\hat{T}$ term $T_\tau = \braket{\theta^{\tau+1}|e^{-\delta\tau\hat{T}}|\theta^\tau}$. Using the fact that the eigenstates of $\hat{T}$ are number states $\ket{n^\tau} = \prod_\mathbf{r}\ket{n_\mathbf{r}^\tau}$, we can insert a complete set of number states for each timeslice and obtain

\begin{align}
T_\tau &= \prod_{\mathbf{r}}\sum_{n_\mathbf{r}^\tau=-\infty}^\infty\braket{\theta_\mathbf{r}^{\tau+1}|e^{-\delta\tau\hat{T}}|n_\mathbf{r}^\tau}\braket{n_\mathbf{r}^\tau|\theta_\mathbf{r}^\tau}\\
&= \prod_{\mathbf{r}}\sum_{n_\mathbf{r}^\tau=-\infty}^\infty e^{-\delta\tau \frac{1}{2}E_C{n_\mathbf{r}^\tau}^2}e^{in_\mathbf{r}^\tau\left(\theta_\mathbf{r}^\tau-\theta_\mathbf{r}^{\tau+1}\right)}
\label{eq:kinetic}
\end{align}
where in the last line we have used the fact that the overlap of phase and number states takes the form
$\braket{\theta_\mathbf{r}^{\tau^\prime}|n_\mathbf{r}^\tau} = e^{-in_\mathbf{r}^\tau\theta_\mathbf{r}^{\tau^\prime}}$ up to a normalization factor that can be absorbed into $\mathcal{D}\theta$.

The sum in $T_\tau$ now has the form $S\left(\phi\right) = \sum_{n=-\infty}^\infty e^{-\frac{1}{2\alpha}n^2}e^{in\phi}$
which can be rewritten using the Poisson summation formula in terms of the summand's Fourier transform as $S\left(\phi\right) = \sum_{k=-\infty}^\infty \sqrt{2\pi\alpha}e^{-\frac{\alpha}{2}\left(\phi-2\pi k\right)^2}$.

Note that in this relation $\alpha = (\delta\tau E_C)^{-1}$. In the limit of large $\alpha$ (small $\delta\tau$) this periodic sum of Gaussians serves as the Villain approximation\cite{villain1976} of the function $e^{\alpha\cos\phi}$

\begin{equation}
S\left(\phi\right) \overset{\alpha\rightarrow\infty}{\sim} e^{-\alpha}e^{\alpha\cos\phi}
\end{equation}
which allows $T_\tau$ to take the desired form (the constant prefactor can be absorbed into $\mathcal{D}\theta$)

\begin{align}
T_\tau \approx e^{(\delta\tau E_C)^{-1}\sum_\mathbf{r}\cos\left(\theta_\mathbf{r}^\tau-\theta_\mathbf{r}^{\tau+1}\right)}.
\end{align}
This leads to the fully mapped classical partition function

\begin{equation}
\label{eq:part5}
Z \approx \int\mathcal{D}\theta e^{-S_{\mathrm{XY}}}
\end{equation}
This is the partition function for a classical anisotropic XY model in ($d+1$)-dimensions with action

\begin{equation}
\label{eq:sxy}
S_{\mathrm{XY}} = -K_\tau\sum_{\mathbf{r},\tau}\cos\left(\theta_\mathbf{r}^\tau-\theta_\mathbf{r}^{\tau+1}\right)-K_\mathbf{r}\sum_{\braket{\mathbf{r}\mathbf{r'}},\tau}\cos\left(\theta_\mathbf{r}^\tau-\theta_{\mathbf{r'}}^\tau\right)
\end{equation}
where the couplings are given by

\begin{equation}
\label{eq:coup}
K_\tau \equiv (\delta\tau E_C)^{-1} \qquad K_\mathbf{r} \equiv \delta\tau E_J.
\end{equation}
We simulate this model with Monte Carlo using an efficient Wolff algorithm where we tune $g = E_C/E_J$ by altering the couplings $K_\tau$ and $K_\mathbf{r}$ and tune temperature by changing the number of timeslices $M$. All global simulations are run on a $64\times64$ lattice at temperature $T = 0.15625 E_J$ corresponding to $M = 64$. All local simulations are run on a $24\times24$ lattice at the same temperature and at bond disorder $p = 0.1$. 

\subsection*{Appendix A2: Mapping to Integer Current Model}
To map to the ICM, we start with the partition function~\eqref{eq:part5} using the form of $S_{\mathrm{XY}}$ given by~\eqref{eq:sxy}. The goal is to sum over all $\theta$ configurations and rewrite $Z$ as a sum over classical integer current configurations $j$. To do this, we use use the Fourier series expansion of $e^{\alpha\cos\phi}$

\begin{equation}
    e^{\alpha\cos\phi} = \sum_{j=-\infty}^\infty I_j(\alpha)e^{ij\phi}
\end{equation}
where $I_j$ is the modified Bessel function of order $j$. Using the shorthand $\partial_b\theta_\mathbf{r}^{\tau}$ to denote a phase difference along a generic spacetime bond $b \in (\hat{\tau},\hat{x},\hat{y})$, we get

\begin{equation}
    e^{K_b \cos (\partial_b\theta_\mathbf{r}^{\tau})} = \sum_{j_{\mathbf{r},b}^\tau=-\infty}^\infty I_{j_{\mathbf{r},b}^\tau}(K_b)e^{ij_{\mathbf{r},b}^\tau\partial_b\theta_\mathbf{r}^{\tau}}
\end{equation}
and inserting into~\eqref{eq:part5} $Z$ becomes

\begin{equation}
    Z \approx \int\mathcal{D}\theta \prod_{\mathbf{r},\tau,b} \sum_{j_{\mathbf{r},b}^\tau=-\infty}^\infty I_{j_{\mathbf{r},b}^\tau}(K_b)e^{ij_{\mathbf{r},b}^\tau\partial_b\theta_\mathbf{r}^{\tau}}.
\end{equation}

To obtain the ICM, we simply have to evaluate the remaining integrals over $\theta$. Note that for each $\theta$ these integrals take the form
\begin{equation}
    \int_0^{2\pi} \frac{d\theta_{\mathbf{r}}^\tau}{2\pi} e^{i \sum_{b\in\pm(\hat{\tau},\hat{x},\hat{y})} j_{\mathbf{r},b}^\tau \theta_{\mathbf{r}}^\tau} = \delta_{\mathbf{\Delta}\cdot\mathbf{j}_\mathbf{r}^\tau, 0}
\end{equation}
where $\mathbf{\Delta}$ is the discrete gradient applied across spacetime. This effectively puts a Kirchhoff's law-like constraint on the currents $j_{\mathbf{r},b}^\tau$; the sum of the spacetime currents $j_{\mathbf{r},b}^\tau$ flowing into a site $(\tau,\mathbf{r})$ must equal the sum of currents flowing out. 

We finally obtain a partition function written in terms of only the $j_{\mathbf{r},b}^\tau$ integer currents
\begin{align}
    Z &\approx \sum_{[j_{\mathbf{r},b}^\tau]}\:'\:\prod_{\mathbf{r},\tau,b} I_{j_{\mathbf{r},b}^\tau}(K_b)\\
    &\approx \sum_{[j_{\mathbf{r},b}^\tau]}\:'\: e^{-S_{\mathrm{ICM}}}
\end{align}
where $\sum_{[j_{\mathbf{r},b}^\tau]}\:'\:$ is a set of sums over all integer currents $j_{\mathbf{r},b}^\tau$ in spacetime subject to the constraint $\mathbf{\Delta}\cdot\mathbf{j}_\mathbf{r}^\tau = 0$ and $S_{\mathrm{ICM}}$ takes the form

\begin{equation}
    S_{\mathrm{ICM}} = -\log\prod_{\mathbf{r},\tau,b} I_{j_{\mathbf{r},b}^\tau}(K_b)
\end{equation}
with corresponding couplings $K_\mathbf{r}$ and $K_\mathbf{\tau}$ given by~\eqref{eq:coup}. $S_{\mathrm{ICM}}$ is a model of integer currents flowing on a spacetime lattice subject to a divergenceless constraint. We perform Monte Carlo simulations on this model using a worm algorithm tuning $g$ and $T$ the same way we do in the XY model. All simulations are performed on a $64\times64$ spatial lattice at varying temperatures. Local simulations are performed on the same lattice size with a spatial bond disorder $p = 0.1$.

\section*{Appendix B: Sum Rules for Maximum Entropy Method}
\begin{figure}[!tb]
\includegraphics[width=0.5\textwidth]{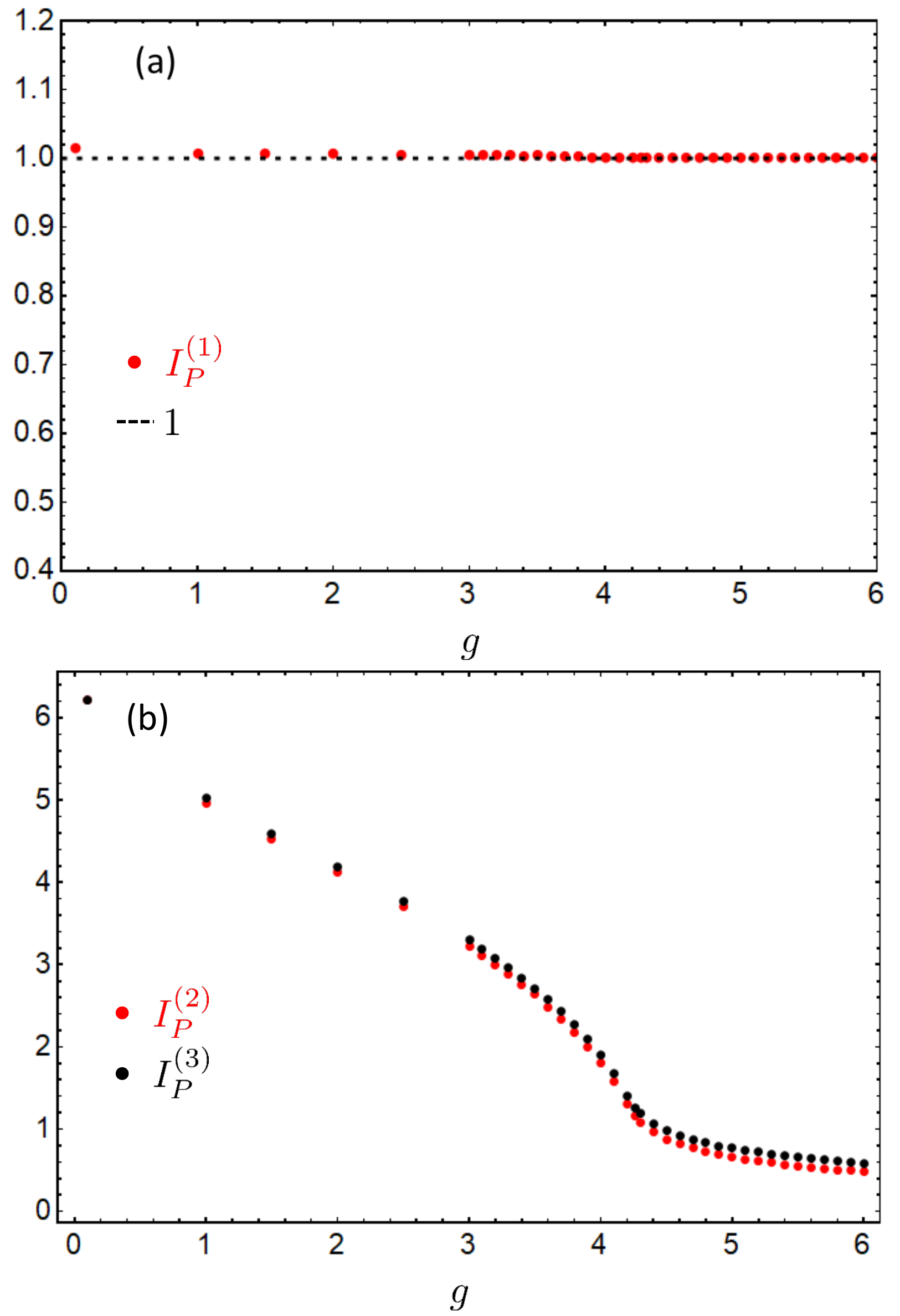}
\caption{We check the accuracy of our analytic continuation procedure using sum rules (a) $I_P^{(1)}$ = 1 and (b) $I_P^{(2)}$ = $I_P^{(3)}$ as described in (\ref{eq:sumrule1}) and (\ref{eq:sumrule23}) respectively. Our data matches the sum rules well.}
\label{sumrules}
\end{figure}
As shown in (\ref{eq:laplace}), the spectral function $P(\omega)$ is related to the Green's function $G(\omega)$ by the inversion of a Laplace transform
\begin{equation}
\label{eq:laplace2}
    G(\tau) = \int_{-\infty}^\infty \frac{d\omega}{\pi} \frac{e^{-\tau\omega}}{1-e^{-\beta\omega}} P(\omega)
\end{equation}
To check the validity of our analytic continuation procedure, we can derive several sum rules from this relation. First, note that $G(\tau = 0) = 1$ must be true. This leads to the first sum rule
\begin{equation}
\label{eq:sumrule1}
    I_P^{(1)} = \int_{-\infty}^\infty \frac{d\omega}{\pi} \frac{1}{1-e^{-\beta\omega}} P(\omega) = G(\tau = 0) = 1.
\end{equation}
The second sum rule can be obtained by integrating both sides of (\ref{eq:laplace2}) with respect to $\tau$ from $0$ to $\beta$
\begin{equation}
\label{eq:sumrule23}
    I_P^{(2)}  = \int_{-\infty}^\infty \frac{d\omega}{\pi} \frac{P(\omega)}{\omega} = \int_{0}^\beta d\tau G(\tau) = I_P^{(3)}.
\end{equation}
We plot the results of the sum rules as a function of $g$ in Fig.~\ref{sumrules} and find our analytically continued data to be in good agreement with them.

\begin{thebibliography}{99}

\bibitem{hebard1990}
A.~F. Hebard and M.~A. Paalanen, Phys. Rev. Lett. {\bf 65}, 927
  (1990).
  
\bibitem{shahar1992}
D.~Shahar and Z.~Ovadyahu, Phys. Rev. B {\bf 46}, 10917 (1992).

\bibitem{goldman1998}
A. Goldman and N. Markovi\'{c}, Physics Today {\bf 51}, 39 (1998).

\bibitem{adams2004}
P.~Adams, Phys. Rev. Lett. {\bf 92}, 067003 (2004).

\bibitem{sambandamurthy2004}
G.~Sambandamurthy, L.~W. Engel, A.~Johansson, and D.~Shahar, Phys. Rev.
  Lett. {\bf 92}, 107005 (2004).
  
\bibitem{steiner2005}
M.~A. Steiner, G.~Boebinger, and A.~Kapitulnik, Phys. Rev. Lett. {\bf
  94}, 107008 (2005).

\bibitem{stewart2007}
M.~D. Stewart, A.~Yin, J.~M. Xu, and J.~M. Valles, Science {\bf 318},
  1273 (2007).
 
\bibitem{gantmakher2010}
V.~F. Gantmakher and V.~T. Dolgopolov, Physics-Uspekhi {\bf 53}, 3
  (2010).
  
\bibitem{sachdev}
S. Sachdev, \textit{Quantum Phase Transitions} (Cambridge, 2011), 2nd ed.

\bibitem{ghosal1998}
A.~Ghosal, M.~Randeria, and N.~Trivedi, Phys. Rev. Lett. {\bf 81},
  3940 (1998).

\bibitem{ghosal2001}
A.~Ghosal, M.~Randeria, and N.~Trivedi, Phys. Rev. B {\bf 65}, 014501
  (2001).

\bibitem{bouadim2011}
K.~Bouadim, Y.~L. Loh, M.~Randeria, and N.~Trivedi, Nature Physics {\bf 7},
  884 (2011).
  
\bibitem{sacepe2008}
B. Sac{\'e}p{\'e}, C. Chapelier, T.~I. Baturina, V.~M. Vinokur, M.~R. Baklanov, and M. Sanquer, Phys. Rev. Lett. {\bf 101}, 157006, (2008).
  
\bibitem{sacepe2011}
B. Sac{\'e}p{\'e}, T. Dubouchet, C. Chapelier, M. Sanquer, M. Ovadia, D. Shahar, M. Feigel'man, and L. Ioffe, Nature Physics {\bf 7}, 239 (2011).

\bibitem{mondal2011}
M. Mondal, A. Kamlapure, M. Chand, G. Saraswat, S. Kumar, J. Jesudasan, L. Benfatto, V. Tripathi, and P. Raychaudhuri, Phys. Rev. Lett. {\bf 106}, 047001 (2011).

\bibitem{sherman2012}
D.~Sherman, G.~Kopnov, D.~Shahar, and A.~Frydman, Phys. Rev. Lett. {\bf 108}, 177006 (2012).

\bibitem{wallin1994}
M. Wallin, E.~S. S{\o}rensen, S.~M. Girvin, and A.~P. Young, Phys. Rev. B {\bf 49}, 12115 (1994).

\bibitem{sondhi1997}
S.~L. Sondhi, S.~M. Girvin, J.~P. Carini, and D. Shahar, Rev. Mod. Phys. {\bf 69}, 315 (1997).

\bibitem{swanson2014}
M. Swanson, Y.~L. Loh, M. Randeria, and N. Trivedi, Phys. Rev. X {\bf 4}, 021007 (2014).

\bibitem{crane2007}
R.~W. Crane, N.~P. Armitage, A. Johansson, G. Sambandamurthy, D. Shahar, and G. Gr\"{u}ner, Phys. Rev. B {\bf 75}, 094506 (2007).

\bibitem{sherman2015}
D. Sherman, U.~S. Pracht, B. Gorshunov, S. Poran, J. Jesudasan, M. Chand, P. Raychaudhuri, M. Swanson, N. Trivedi, A. Auerbach, M. Scheffler, A. Frydman, and M. Dressel, Nature Physics {\bf 11}, 188 (2015).

\bibitem{poran2017}
S. Poran, T. Nguyen-Duc, A. Auerbach, N. Dupuis, A. Frydman, and O. Bourgeois, Nature Communications {\bf 8}, 14464 (2017).

\bibitem{kremen2018}
A. Kremen, H. Khan, Y.~L. Loh, T.~I. Baturina, N. Trivedi, A. Frydman, and B. Kalisky, Nature Physics {\bf 14}, 1205 (2018).

\bibitem{howald2001}
C. Howald, P. Fournier, and A. Kapitulnik, Phys. Rev. B {\bf 64}, 100504(R) (2001).

\bibitem{pan2001}
S.~H. Pan, J.~P. O'Neal, R.~L. Badzey, C. Chamon, H. Ding, J.~R. Engelbrecht, Z. Wang, H. Eisaki, S. Uchida, A.~K. Gupta, K.-W. Ng, E.~W. Hudson, K.~M. Lang, and J.~C. Davis, Nature {\bf 413}, 282 (2001).

\bibitem{lang2002}
K.~M. Lang, V. Madhavan, J.~E. Hoffman, E.~W. Hudson, H. Eisaki, S. Uchida, and J.~C. Davis, Nature {\bf 415}, 412 (2002).

\bibitem{martin2008}
J. Martin, N. Akerman, G. Ulbricht, T. Lohmann, J.H. Smet, K. von Klitzing, and A. Yacoby, Nature Physics {\bf 4}, 144 (2008).

\bibitem{nadj-perge2014}
S. Nadj-Perge, I.~K. Drozdov, J. Li, H. Chen, S. Jeon, J. Seo, A.~H. MacDonald, B.~A. Bernevig, and A. Yazdani, Science {\bf 346}, 602 (2014).

\bibitem{randeria2016}
M.T. Randeria, B.E. Feldman, I.K. Drozdov, and A. Yazdani, Phys. Rev. {\bf B}, 161115(R) (2016).

\bibitem{mooij2006}
J.~E. Mooij and Yu.~V. Nazarov, Nature Physics {\bf 2}, 169 (2006).

\bibitem{wolff1989}
U. Wolff, Phys. Rev. Lett. {\bf 62}, 361 (1989).

\bibitem{prokofev2001}
N. Prokof'ev and B. Svistunov, Phys. Rev. Lett. {\bf 87}, 160601 (2001).

\bibitem{fisher1989}
M.~P.~A. Fisher, P~B. Weichman, G. Grinstein, and D.~S. Fisher, Phys. Rev. B {\bf 40}, 546 (1989).

\bibitem{gubernatis1991}
J.~E. Gubernatis, M.~Jarrell, R.~N. Silver, and D.~S. Sivia,  Phys. Rev. B {\bf 44}, 6011 (1991).
  
\bibitem{sandvik1998}
A.~W. Sandvik, Phys. Rev. B {\bf 57}, 10287 (1998).

\bibitem{kubo2003}
R. Kubo, M. Toda, and N. Hashitsume, \textit{Statistical Physics II: Nonequilibrium Statistical Mechanics} (Springer, 2003), 2nd ed.

\bibitem{suzuki1993}
M. Suzuki, Physica A {\bf 194}, 432 (1993).

\bibitem{villain1976}
J. Villain, J. Phys. France {\bf 36}, 581 (1976).

\end{thebibliography}
\end{document}